\documentclass[twocolumn,aps,showpacs,prb]{revtex4-1}

\usepackage{amsmath,amssymb,graphics,epsfig,epstopdf,color,multirow,array,verbatim,ulem,braket,tabularx}
\usepackage[colorlinks,linkcolor=blue,citecolor=blue,urlcolor=blue]{hyperref}

\begin{document}

\title{Magnetic interactions in a proposed diluted magnetic semiconductor (Ba$_\text{1-x}$K$_\text{x}$)(Zn$_\text{1-y}$Mn$_\text{y}$)$_\text{2}$P$_\text{2}$}

\author{Huan-Cheng Yang}
\author{Kai Liu}
\email{kliu@ruc.edu.cn}
\author{Zhong-Yi Lu}
\email{zlu@ruc.edu.cn}
\affiliation{Department of Physics and Beijing Key Laboratory of Opto-electronic Functional Materials $\&$ Micro-nano Devices, Renmin University of China, Beijing 100872, China}

\begin{abstract}
 By using first-principles electronic structure calculations, we have studied the magnetic interactions in a proposed BaZn$_2$P$_2$-based diluted magnetic semiconductor (DMS). For a typical compound Ba(Zn$_{0.944}$Mn$_{0.056}$)$_2$P$_2$ with only spin doping, due to the superexchange interaction between Mn atoms and the lack of itinerant carriers, the short-range antiferromagnetic coupling dominates. Partially substituting K atoms for Ba atoms, which introduces itinerant hole carriers into the $p$ orbitals of P atoms so as to link distant Mn moments with the spin-polarized hole carriers via the $p$-$d$ hybridization between P and Mn atoms, is very crucial for the appearance of ferromagnetism in the compound. Furthermore, applying hydrostatic pressure first enhances and then decreases the ferromagnetic coupling in (Ba$_{0.75}$K$_{0.25}$)(Zn$_{0.944}$Mn$_{0.056}$)$_2$P$_2$ at a turning point around 15 GPa, which results from the combined effects of the pressure-induced variations of electron delocalization and $p$-$d$ hybridization. Compared with the BaZn$_2$As$_2$-based DMS, the substitution of P for As can modulate the magnetic coupling effectively. Both the results for BaZn$_2$P$_2$-based and BaZn$_2$As$_2$-based DMSs demonstrate that the robust antiferromagnetic (AFM) coupling between the nearest Mn-Mn pairs bridged by anions is harmful to improving the performance of this II-II-V based DMS materials.
\end{abstract}

%\keywords{BaZn$_2$P$_2$-based DMS, $p$-$d$ hybridization, magnetic coupling, first-principles calculations}


\date{\today} \maketitle

\section{Introduction}
The diluted magnetic semiconductors (DMSs), which are semiconductors doped with magnetic impurities, have attracted extensive attention due to their potential applications in spintronic devices as well as their fundamental scientific values~\cite{Science_I, Science_II, Nature_I, NatureMaterials_I, RevModPhys_I, ZJH, ZJH_I}. Experimentally, the synthesis of materials combining the semiconducting behavior with the robust ferromagnetism has long been a dream of material physicists and much effort has actually been devoted to searching for such materials with Curie temperature ($T_{c}$) above room temperature~\cite{NatureMaterials_II, PRL_I, APL_I, APL_II, PRB_I, Nano_I}. Theoretically, the focus is placed on the understanding of magnetic mechanism in order to provide suggestions for exploring high-$T_{c}$ materials~\cite{JAP_I, PRB_II, PRB_III, RevModPhys_II, RevModPhys_III}. However, despite several decades of intensive work, the preparation of materials with practical feasibility is still a challenge and the complexity of real materials often hinders a clear theoretical understanding.

The exploration of DMS materials mainly involves the semiconductors doped with Mn atom due to its large local magnetic moment. In the early days, the Mn doped (A$_{1-x}^\text{II}$Mn$_{x}$)B$^\text{VI}$ compounds were often chosen as prototypical systems~\cite{JAP_I}, as the isovalent (Mn$^{2+}$, A$^\text{II}$) substitution would make them easy to achieve a high solubility of Mn. However, the initial charge doping in the II-VI based DMSs, no matter whether $n$-type or $p$-type, was difficult, until the later progress in charge doping leads to the emergence of their ferromagnetism~\cite{APL_IV, PRL_II, PRB_IV, Handbook_I}. Subsequently, the III-V based DMSs were widely studied due to their compatibility with the present-day electronic materials. Among them, the (Ga, Mn)As has achieved a $T_{c}$ around 200 K~\cite{Nano_I}.
Nevertheless, due to the hetero-valent (Mn$^{2+}$, Ga$^{3+}$) substitution, the equilibrium solubility of Mn in the III-V compounds is very small, such DMS samples are thus often prepared as films by non-equilibrium epitaxial growth technique~\cite{Handbook_I}. Moreover, the hetero-valent (Mn$^{2+}$, Ga$^{3+}$) substitutions introduce spins and holes simultaneously, which makes the independent control of spin- and charge-doping impossible~\cite{Handbook_I}. Recently, the I-II-V compound Li(Zn, Mn)As~\cite{NatureCommunications_Deng}, the II-II-V compound (Ba, K)(Zn, Mn)$_{2}$As$_{2}$~\cite{NatureCommunications_I}, and the III-VI-II-V compound (La, Ba)O(Zn, Mn)As~\cite{PRB_Ning}, which are respectively isostructural to the well-known '111', '122', and '1111' iron-based superconductors LiFeAs~\cite{SSC}, BaFe$_{2}$As$_{2}$~\cite{PRL_III}, and LaOFeAs~\cite{JACS}, have been reported as new types of DMS materials. The $T_{c}$ of (Ba$_{0.7}$K$_{0.3}$)(Zn$_{0.85}$Mn$_{0.15}$)$_{2}$As$_{2}$ even reaches $\sim$230 K~\cite{CSB}, higher than the record value achieved in (Ga, Mn)As. More importantly, in this new '122' type DMS material, the spin doping by isovalent (Mn$^{2+}$, Zn$^{2+}$) substitution and the charge doping by hetero-valent (Ba$^{2+}$, K$^{+}$) substitution decouple with each other, thus providing us an unique opportunity to study the magnetic mechanism in DMS.

From the theoretical standpoint, previous works on the (Ba$_\text{1-x}$K$_\text{x}$)(Zn$_\text{1-y}$Mn$_\text{y}$)$_2$As$_2$ compound~\cite{PRB_V, ComputMaterSicence_I, SolidStatePhysics_I, PRB_VI} propose the existence of both short-range antiferromagnetic (AFM) interactions via superexchange and long-range ferromagnetic (FM) interactions mediated by itinerant holes. Thus, the nearest-neighbour Mn atoms often take the AFM coupling, yielding a reduction of the mean magnetization of all Mn atoms compared with the local moment of Mn$^{2+}$. Furthermore, the analysis based on density functional theory (DFT) calculations by Mazin $et$. $al$. gave an excellent agreement of magnetization between their calculation results and experiment data, which demonstrates that the DFT can play an important role in understanding this recently discovered II-II-V type DMS materials~\cite{PRB_V}.

 As a counterpart of (Ba$_\text{1-x}$K$_\text{x}$)(Zn$_\text{1-y}$Mn$_\text{y}$)$_2$As$_2$, (Ba$_\text{1-x}$K$_\text{x}$)(Zn$_\text{1-y}$Mn$_\text{y}$)$_2$P$_2$ (with P atom substituting the same group As) is expected to be a similar DMS. More importantly, the substitution of As with P would introduce some changes in the magnetic properties of (Ba$_\text{1-x}$K$_\text{x}$)(Zn$_\text{1-y}$Mn$_\text{y}$)$_2$P$_2$, the study on which may enable us a complete understanding on the exchange interactions in this prototypical DMS materials and may provide guidance for searching more feasible DMS candidates. Here, we have carried out systematic investigations on the proposed (Ba$_\text{1-x}$K$_\text{x}$)(Zn$_\text{1-y}$Mn$_\text{y}$)$_2$P$_2$ compound to explore the magnetic interactions in it.

\section{Computational details}
\label{sec:Method}

First-principles electronic structure calculations were performed by using the projector augmented wave (PAW) method~\cite{PAW_I, PAW_II} as implemented in the Vienna Ab initio Simulation Package~\cite{VASP_I, VASP_II, VASP_III}. The generalized gradient approximation (GGA) of Perdew-Burke-Ernzerhof (PBE) type was employed for the exchange-correlation functional~\cite{PBE}. The kinetic energy cutoff of the plane-wave basis was set to be 400 eV. A fully variable-cell relaxation of BaZn$_2$P$_2$ unit cell with 10 atoms was first carried out to obtain the equilibrium lattice parameters under different pressures. The criteria for force convergence on all atoms was 0.01 eV/{\AA}. Then the properties of the BaZn$_2$P$_2$ parent compound were studied. By tripling these relaxed unit cells along both $a$ and $b$ directions, we obtained the expanded supercells containing 90 atoms for later studies on the effects of spin- and charge- doping.

\begin{figure}[!t]
\centering
\includegraphics[width=1.0\columnwidth]{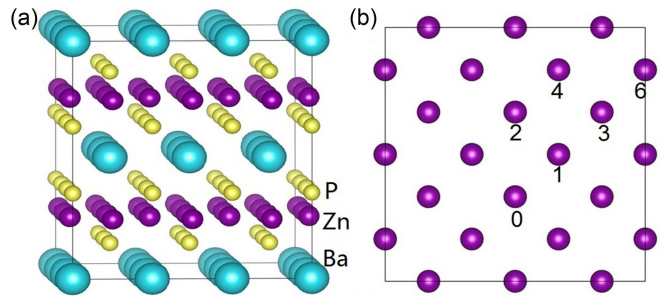}
\caption{\small(Color online) (a) The schematic diagram of the BaZn$_\text{2}$P$_\text{2}$ supercell. (b) The substitution sites of Mn atoms in the same Zn layer. Two Mn atoms are arranged as 0-1, 0-2, 0-3, 0-4, and 0-6 pairs, while the integers represent the sequences of Mn neighbours.}
\label{fig:Structure}
\end{figure}

The supercell we used is schematically shown in Figure~\ref{fig:Structure}(a). It contains two Zn layers stacked along the [001] direction. For the spin doping, we used two Mn atoms to substitute two Zn atoms respectively, which introduces a doping concentration of 5.6$\%$. The two Mn atoms can locate either in the same Zn layer or in different Zn layers. In the same Zn layer, they can locate at sites 0, 1, 2, 3, 4, or 6 [Figure~\ref{fig:Structure}(b)] to form Mn-Mn pairs denoted as 0-1, 0-2, 0-3, 0-4, and 0-6, corresponding to the first, second, third, fourth, and sixth neighbors, respectively. If the corresponding substitution sites of Mn in the other Zn layer are denoted by $0^{'}$, $1^{'}$, $2^{'}$, $3^{'}$, $4^{'}$, and $6^{'}$, the two Mn atoms can also form Mn-Mn pairs named 0-$0^{'}$, 0-$1^{'}$, 0-$2^{'}$, 0-$3^{'}$, 0-$4^{'}$, and 0-$6^{'}$ respectively. For all supercells of Ba(Zn$_\text{0.944}$Mn$_\text{0.056}$)$_2$P$_2$, we relaxed the internal atomic positions only. Regarding the charge doping, we employed the virtual crystal approximation (VCA), in which the Ba sites were composed of 75$\%$ Ba$^\text{2+}$ and 25$\%$ K$^\text{+}$. Both the concentrations of spin doping 5.6$\%$ and hole doping 25$\%$ are typical experimental values for the '122' type DMSs~\cite{NatureCommunications_I}. The electronic correlation effect among Mn 3$d$ electrons was incorporated by the DFT+U formalism according to the method of Dudarev $et$ $al$. (effective U)~\cite{DFTU-II}. The value of effective U was set to 3.0 eV, which had been used and discussed carefully in the previous work for compounds Li(ZnMn)As~\cite{SSC_177,PRL_98} and (BaK)(ZnMn)$_2$As$_2$~\cite{ComputMaterSicence_I}.

It is known that for the BaZn$_2$As$_2$, namely the counterpart of BaZn$_2$P$_2$, there are two crystalline phases: the low-temperature orthorhombic phase ($\alpha$-BaZn$_2$As$_2$ with space group $Pnma$) and the high-temperature tetragonal phase ($\beta$-BaZn$_2$As$_2$ with space group $I4/mmm$)~\cite{ThCrSi}. Experimentally, under low temperature, the stable $\beta$-BaZn$_2$As$_2$ at ambient condition can be obtained by the rapid quenching method~\cite{JACS_I}. Moreover, 10$\%$ of K or Mn doping can stabilize the tetragonal $\beta$-BaZn$_2$As$_2$ down to 3.5 K~\cite{NatureCommunications_I}. Here, we take the tetragonal $\beta$-BaZn$_2$P$_2$ to perform the calculations. At ambient pressure, the calculated equilibrium lattice constants of the BaZn$_2$P$_2$ (BaZn$_2$As$_2$) tetragonal unit cell are a = 4.039 \AA (4.156 \AA) and c = 13.280 \AA (13.641 \AA), which are in good accordance with the experimental values a = 4.019 \AA (4.12 \AA) and 13.228 \AA (13.58 \AA)~\cite{NatureCommunications_I, ThCrSi, BZNP}. Actually, according to our calculations, the energy of $\beta$-BaZn$_2$P$_2$ ($\beta$-BaZn$_2$As$_2$) is just 0.138 (0.148) eV per formula unit higher than that of $\alpha$-BaZn$_2$P$_2$ ($\alpha$-BaZn$_2$As$_2$).

\section{Results and analysis}
\label{sec:Results}

\subsection{Carrier-mediated ferromagnetism}

The classical Heisenberg model reading
\begin{equation}
H_{mag}=-\sum_{i\neq{j}}J_{ij}\textbf{e}_{i}\cdot{\mathbf{e}_{j}}
\end{equation}
has been employed to study the magnetic interactions. Here $J_{ij}$  is the exchange integral parameter and \textbf{e}$_{i}$ is the unit vector in the direction of the spin \textbf{S}$_{i}$ on site $i$ with the moment $M$ (\textbf{e}$_{i}$=\textbf{S}$_{i}$/$M$). The calculated total enthalpies for the ferromagnetic and antiferromagnetic configurations are denoted as $E_{FM}$ and $E_{AFM}$, respectively. Then the exchange energy $E_{mag}$ of the Mn-Mn pairs can be derived from their enthalpy difference as $E_{mag}$=$\triangle$$E$/2=($E_{AFM}$-$E_{FM}$)/2.

\begin{figure}[!t]
\centering
\includegraphics[width=1.0\columnwidth]{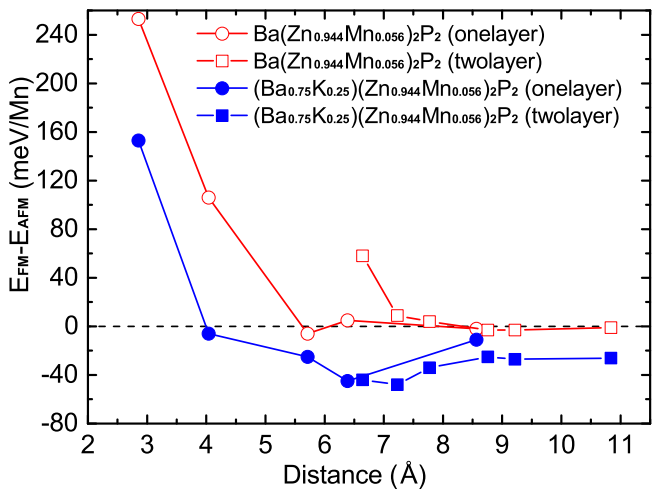}
\caption{\small(Color online) Energy differences per Mn between the FM and AFM states for Ba(Zn$_\text{0.944}$Mn$_\text{0.056}$)$_\text{2}$P$_\text{2}$ and (Ba$_\text{0.75}$K$_\text{0.25}$)(Zn$_\text{0.944}$Mn$_\text{0.056}$)$_\text{2}$P$_\text{2}$ as functions of Mn-Mn distances. The data points denote different Mn-Mn pairs in the supercell. The legend 'onelayer' means that the Mn-Mn pairs are in the same Zn layer while 'twolayer' in different Zn layers. The circles correspond to the 0-1, 0-2, 0-3, 0-4, and 0-6 configurations with increasing Mn-Mn distances while the squares to the 0-$0^{'}$, 0-$1^{'}$, 0-$2^{'}$, 0-$3^{'}$, 0-$4^{'}$, and 0-$6^{'}$ ones.}
\label{fig:Energy-difference I}
\end{figure}

The energy differences (enthalpy differences at finite pressure) between the FM and AFM couplings for all Mn-Mn pairs are shown in Figure~\ref{fig:Energy-difference I}. The data can be classified into two groups: one represents the situation with only spin doping (data in red color), while the other denotes the situation with both spin and hole dopings (data in blue color).  For the former, there are three Mn-Mn pairs, namely 0-1, 0-2, and 0-$0^{'}$, showing robust antiferromagnetism, while the energy differences for the other Mn-Mn pairs are negligible. Thus, the dominant magnetic exchange interactions in Ba(Zn$_\text{0.944}$Mn$_\text{0.056}$)$_\text{2}$P$_\text{2}$ are short-range antiferromagnetic interactions. When it comes to the situation with both spin and hole dopings in (Ba$_\text{0.75}$K$_\text{0.25}$)(Zn$_\text{0.944}$Mn$_\text{0.056}$)$_\text{2}$P$_\text{2}$, the Mn-Mn couplings change significantly. The most noteworthy feature is that all the former weakly coupled Mn-Mn pairs and the 0-$0^{'}$ Mn-Mn pair take ferromagnetic coupling now, while only the 0-1 Mn-Mn pair still holds antiferromagnetic coupling, yet with much weakened strength. Thus, when doped with holes, the long-range FM couplings dominate the magnetic properties of the (Ba$_\text{0.75}$K$_\text{0.25}$)(Zn$_\text{0.944}$Mn$_\text{0.056}$)$_\text{2}$P$_\text{2}$ compound, which demonstrates that the hole carriers indeed mediate the magnetic interactions and play an important role in inducing the ferromagnetism.

\begin{figure}[!t]
\centering
\includegraphics[width=0.75\columnwidth]{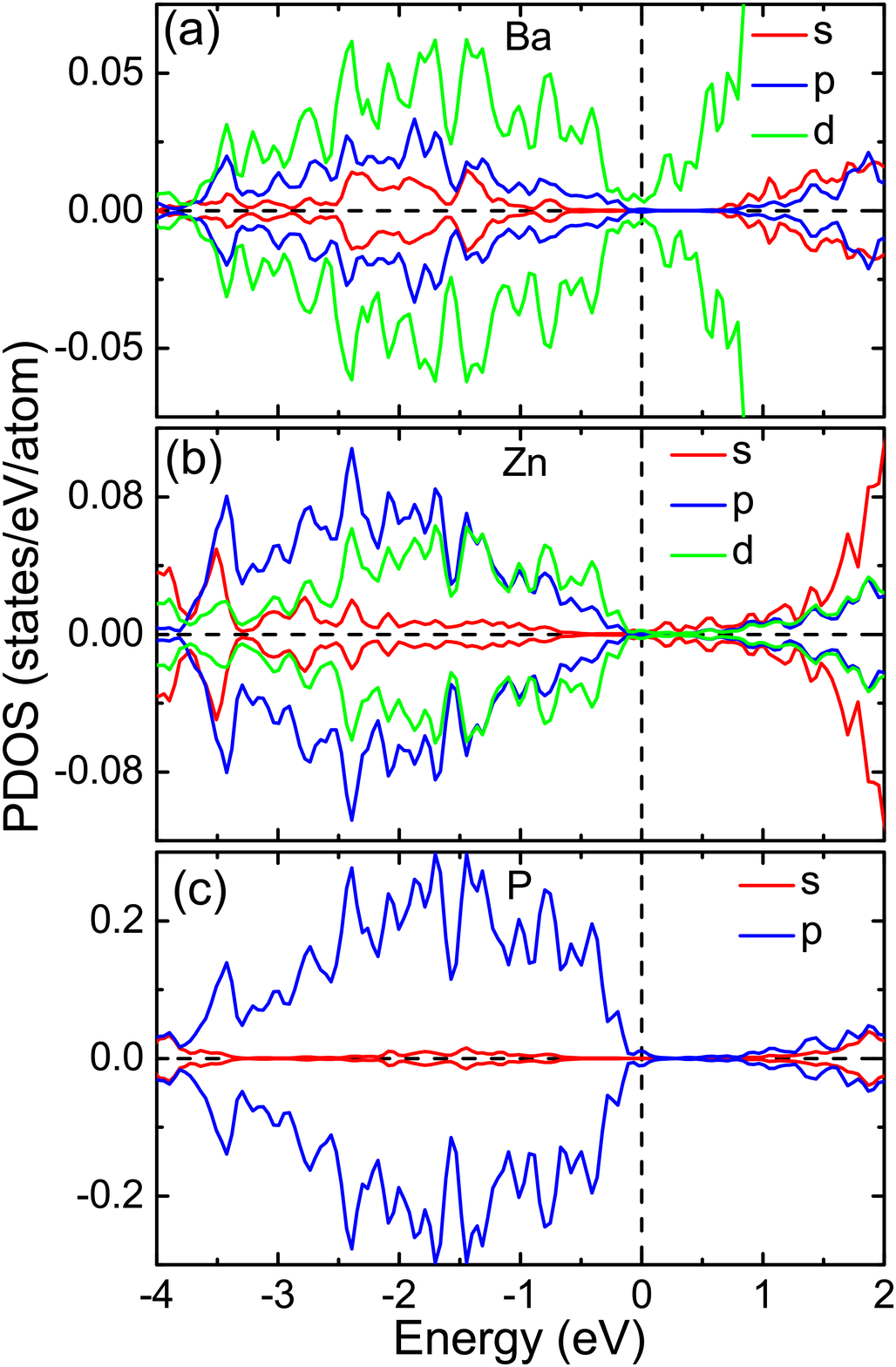}
\caption{\small(Color online) Partial density of states (PDOS) of the parent compound BaZn$_\text{2}$P$_\text{2}$ at 0 GPa. The up and down parts of each panel correspond to the spin-up and spin-down channels, respectively.}
\label{fig:PDOS-BaZnP}
\end{figure}

\subsection{Exchange mechanism}

\begin{figure}[!t]
\centering
\includegraphics[width=1.0\columnwidth,clip=true]{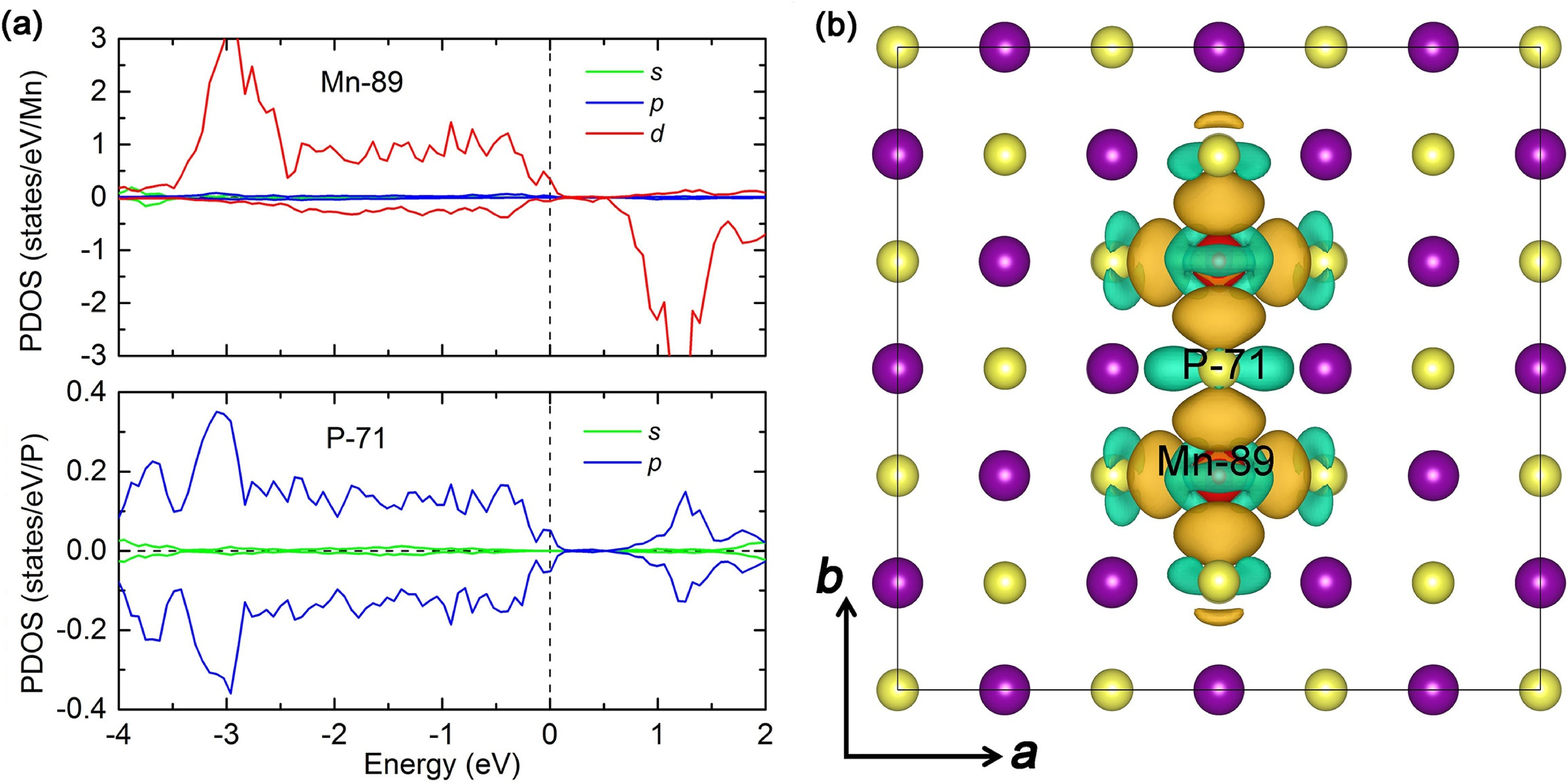}
\caption{\small(Color online) For the 0-2 configuration of Ba(Zn$_\text{0.944}$Mn$_\text{0.056}$)$_\text{2}$P$_\text{2}$ at 0 GPa with antiferromagnetically coupled Mn-Mn pair in the supercell: (a) Respective partial density of states of Mn and P atoms. The up and down parts of each panel correspond to the spin-up and spin-down channels, respectively. The digits are the atom labels in the supercell. (b) Top view of the differential charge densities. Ba atoms are omitted for clarity. The orange and cyan isosurfaces show the electron accumulation and loss regions, respectively. }
\label{fig:PDOS-0-2}
\end{figure}

\begin{figure}[!t]
\centering
\includegraphics[width=1.0\columnwidth,clip=true]{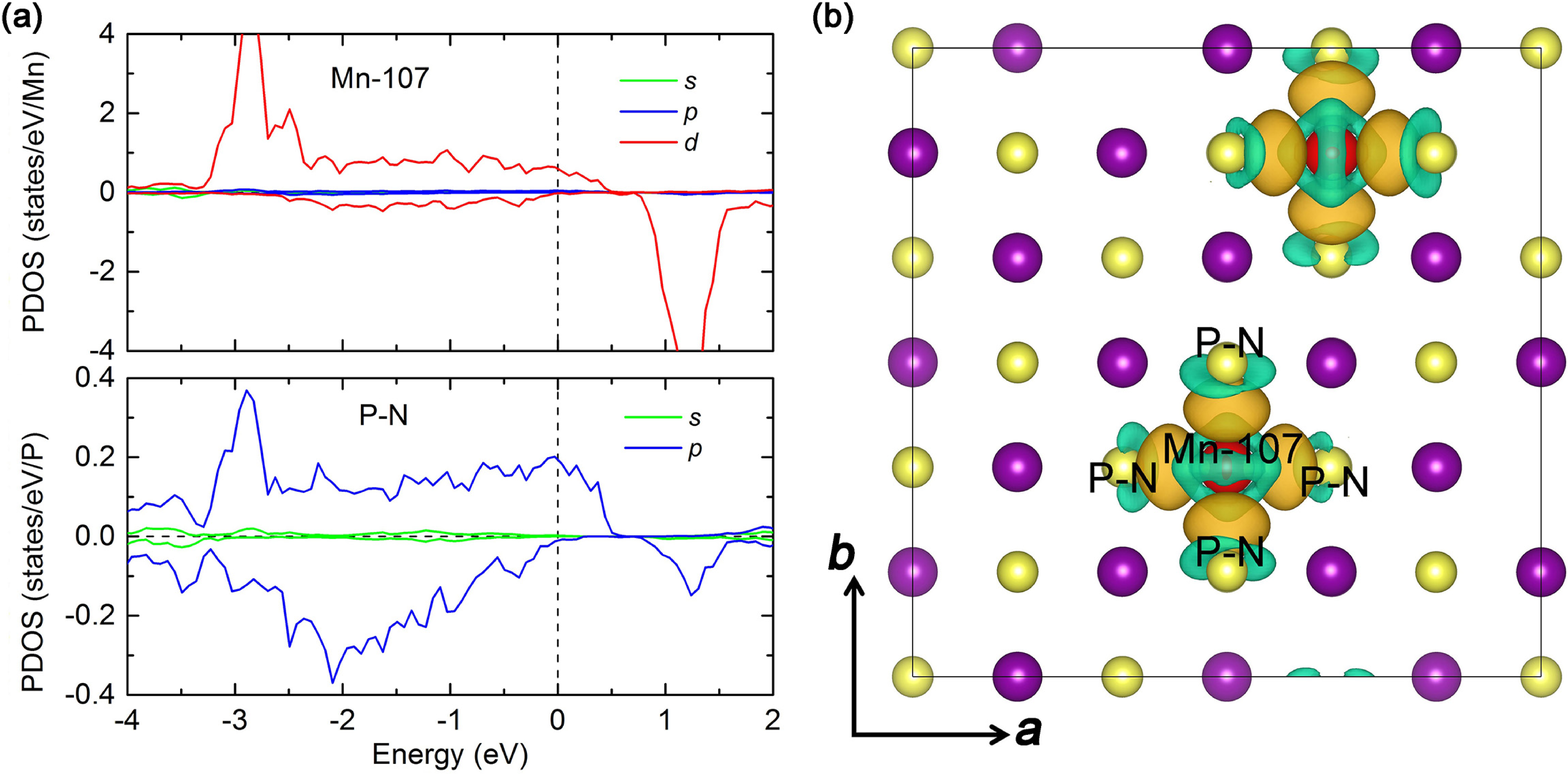}
\caption{\small(Color online) For the 0-4 configuration of (Ba$_\text{0.75}$K$_\text{0.25}$)(Zn$_\text{0.944}$Mn$_\text{0.056}$)$_\text{2}$P$_\text{2}$ at 0 GPa with ferromagnetically coupled Mn-Mn pair in the supercell: (a) Respective partial density of states of Mn and P atoms. The up and down parts of each panel correspond to the spin-up and spin-down channels, respectively. The digits are the atom labels in the supercell. The 'P-N' means the nearest-neighboring P atoms to the Mn atom. (b) Top view of the differential charge densities. Ba atoms are omitted for clarity. The orange and cyan isosurfaces show the electron accumulation and loss regions, respectively. }
\label{fig:PDOS-0-4}
\end{figure}

As presented above, the short-range antiferromagnetic and the long-range ferromagnetic interactions dominate the Mn-Mn couplings in Ba(Zn$_\text{0.944}$Mn$_\text{0.056}$)$_\text{2}$P$_\text{2}$ and (Ba$_\text{0.75}$K$_\text{0.25}$)(Zn$_\text{0.944}$Mn$_\text{0.056}$)$_\text{2}$P$_\text{2}$ compounds, respectively. Here, we show that the mechanism of magnetic interactions in these two BaZn$_\text{2}$P$_\text{2}$-based compounds can be resolved from the electronic structure calculations. For the parent compound BaZn$_\text{2}$P$_\text{2}$, the partial density of states (PDOS) of all three atomic species show reduction around the Fermi level (Figure~\ref{fig:PDOS-BaZnP}), demonstrating a semiconducting behavior. In comparison, for its counterpart $\beta$-BaZn$_2$As$_2$, our calculations give a vanishing gap, namely there being tiny density of states at the Fermi level, which agrees well with previous GGA-type calculations~\cite{ComputMaterSicence_I, SolidStatePhysics_I, JAC_583}. However, the experiment data show a gap of 0.23 eV for $\beta$-BaZn$_2$As$_2$~\cite{JACS_I}. Taking into account their similarity, we speculate, in the absence of experimental results, that BaZn$_2$P$_2$ has a small gap. For these two doped compounds [Figs.~\ref{fig:PDOS-0-2}(a) and ~\ref{fig:PDOS-0-4}(a)], on one hand, the $p$ orbitals of P atoms show a  broad continuum from -4 to 0 eV. Meanwhile, the occupied $d$ orbitals of Mn atoms show main peaks at about -3 eV and delocalize in a range from -3.5 eV to the Fermi level. On the other hand, the Mn$^{2+}$ ions locate in the tetrahedral crystal field created by P atoms and the differential charge densities (definition in Ref.\cite{TuZhu_I}) as shown in Figs.~\ref{fig:PDOS-0-2}(b) and ~\ref{fig:PDOS-0-4}(b) demonstrate obvious electron accumulation between Mn and P atoms. Thus, we suggest that strong $p$-$d$ hybridization between P and Mn atoms is an important characteristic of both doped compounds.

In the following, we demonstrate that the $p$-$d$ hybridization between Mn atom and its neighboring P atoms is a prerequisite in determining the magnetic coupling between the Mn-Mn pairs. On one hand, the $p$-$d$ hybridization can delocalize the $d$ orbitals of Mn atoms and reduce the kinetic energy of the system. When the compound has only spin doping (Figure~\ref{fig:PDOS-0-2}), there is negligible density of states at the Fermi level, namely rare hole carries. If two Mn atoms share a bridging P atom, the antiferromagnetic coupling between these two Mn atoms results from the $p$-$d$ hybridization induced kinetic energy reduction by Pauli exclusion principle, which is considered as AFM superexchange~\cite{PhysicReview_I}. Thus, the 0-1 and 0-2 Mn-Mn pairs [Figure~\ref{fig:Structure}(b)] take the AFM coupling in order to reduce the energy of the compound (Figure~\ref{fig:Energy-difference I}). When there is no bridging P atom between two Mn atoms, each Mn atom hybridizes with its neighboring P atoms respectively, which has almost no requirement for the relative alignment of two Mn spins, yielding a negligible magnetic coupling between them. On the other hand, when doping hole carriers into the $p$ orbitals of P atoms, the strong $p$-$d$ hybridization between Mn and its neighbouring P atoms leads to an itinerant spin-polarized Fermi sea. Namely, the spin-up $p$ orbitals of the P atoms neighboring to spin-up Mn atoms shift to a higher energy while the spin-down $p$ orbitals of P atoms to a lower energy (Figure~\ref{fig:PDOS-0-4}), then the spin-down $p$ states are occupied more than the spin-up $p$ ones, resulting in a spin-polarized Fermi sea. Consequently, this leads to an AFM coupling between the Mn and P atoms, which is considered as Zener's $p$-$d$ exchange~\cite{PhysicReview_II, PRB_VII}. Then, the other Mn atom adjusts its spin polarization to align oppositely to the polarized Fermi sea. Eventually, the two distant Mn atoms, i.e., Mn-Mn pairs without bridging P atoms, take an effective ferromagnetic coupling. This Mn-Mn coupling contains two key points: strong $p$-$d$ hybridization and robust transmission by spin-polarized Fermi sea.
%For the 0-1 and 0-2 Mn-Mn pairs, both the antiferromagnetic superexchange and effective ferromagnetic exchange work together, resulting in a significant weaken of the AFM coupling.

\subsection{Pressure and correlation effects}

\begin{figure*}[!t]
\centering
\includegraphics[width=0.8\textwidth]{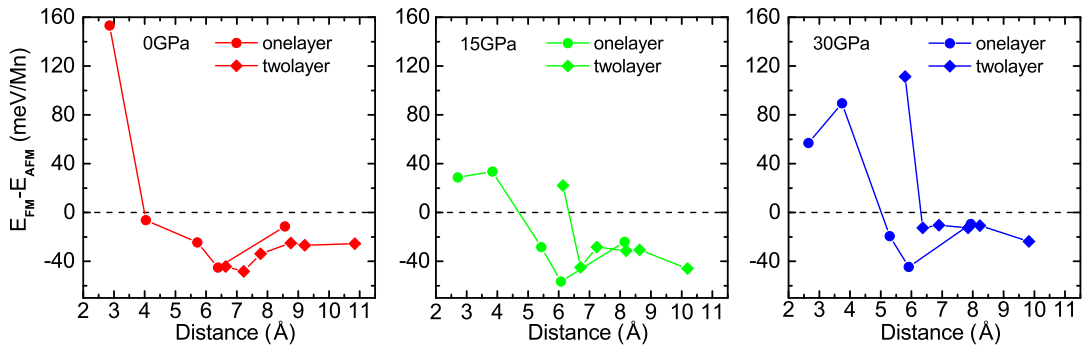}
\caption{\small(Color online) Enthalpy differences per Mn between the FM and AFM states for (Ba$_\text{0.75}$K$_\text{0.25}$)(Zn$_\text{0.944}$Mn$_\text{0.056}$)$_\text{2}$P$_\text{2}$ as a function of Mn-Mn distances at (a) 0 GPa, (b) 15 GPa, and (c) 30 GPa, respectively. The data points denote the different Mn-Mn pairs in the supercell. The legend 'onelayer' means that the Mn-Mn pairs are in the same Zn layer while the 'twolayer' in different Zn layers. The solid circles correspond to the 0-1, 0-2, 0-3, 0-4, and 0-6 configurations with increasing Mn-Mn distances while the solid squares to the 0-$0^{'}$, 0-$1^{'}$, 0-$2^{'}$, 0-$3^{'}$, 0-$4^{'}$, and 0-$6^{'}$ ones.}
\label{fig:Energy-difference II}
\end{figure*}

For this new '122' type DMS, recent X-ray spectroscopy experiments have found that the applied pressure can induce the band broadening of As $p$ orbitals and suppress the exchange interactions~\cite{PRB_Sun, PRB_SunI}. Hence, we study the pressure effect on the magnetic couplings for different Mn-Mn pairs. Once the pressures are applied on (Ba$_\text{0.75}$K$_\text{0.25}$)(Zn$_\text{0.944}$Mn$_\text{0.056}$)$_\text{2}$P$_\text{2}$, both the 0-2 and 0-0' configurations show continuous enhancement of the AFM coupling up to 30 GPa (Figure~\ref{fig:Energy-difference II}). This can be explained by the enhancement of superexchange induced by more orbital overlaps under pressure. Especially for the 0-$0^{'}$ configuration, the reduction of lattice $c$ under pressure may strengthen the overlap of the $p_z$ orbitals between interlayer P atoms, eventually leading to the enhanced AFM coupling. In contrast, the FM coupling for the 0-3, 0-4, 0-6, 0-$3^{'}$, 0-$4^{'}$, and 0-$6^{'}$ Mn-Mn pairs achieve the maxima at 15 GPa. This behavior shows the competition between the two key points on the Mn-Mn effective FM coupling mentioned above. The pressure can strengthen the $p$-$d$ hybridization, meanwhile it can also weaken the spin polarization of the hole carries, as indicated by the reduction of the average spin-polarization on P atoms. The competition between these two aspects leads the nonlinear dependence of the effective FM coupling on pressure.

\begin{table*}[!t]
\caption{\small\label{tab:I}Energy differences per Mn (E$_{mag}$=$\Delta$E/2=($E_{FM}$-$E_{AFM}$)/2, in unit of meV) between the FM and AFM states of Mn-Mn configurations 0-1, 0-2, 0-3, 0-4, 0-6, 0-$0^{'}$, 0-$1^{'}$, 0-$2^{'}$, 0-$3^{'}$, 0-$4^{'}$, and 0-$6^{'}$ at 0 GPa for the (Ba$_\text{0.75}$K$_\text{0.25}$)(Zn$_\text{0.944}$Mn$_\text{0.056}$)$_\text{2}$P$_\text{2}$ (BaZnP-based) and (Ba$_\text{0.75}$K$_\text{0.25}$)(Zn$_\text{0.944}$Mn$_\text{0.056}$)$_\text{2}$As$_\text{2}$ (BaZnAs-based) compounds, respectively.}
\renewcommand\arraystretch{1.4}
\begin{center}
\begin{tabular*}{0.95\textwidth}{@{\extracolsep{\fill}}cccccccccccc}
\hline\hline
    &      0-1&   0-2&   0-3&   0-4&   0-6&   0-$0^{'}$&   0-$1^{'}$&   0-$2^{'}$&   0-$3^{'}$&   0-$4^{'}$&   0-$6^{'}$\\
\hline
  BaZnP-based &    153.2 &   -6.3 &	-24.6 &	   -45.2 &	   -11.5 &	  -44.4 &	 -48.2 &    -33.9 &    -25.0 &	  -26.8 &	   -25.6\\
  BaZnAs-based &    100.8 &	  -17.5 &	 -29.7 &	-43.8 &	   -21.5 &	   -36.8 &	  -40.1 &   -29.0 &	   -23.3 &	  -23.3 &	   -21.0\\
\hline\hline
\end{tabular*}
\end{center}
\end{table*}

Furthermore, we study the correlation effect by using the GGA+U calculations for (Ba$_\text{0.75}$K$_\text{0.25}$)(Zn$_\text{0.944}$Mn$_\text{0.056}$)$_2$P$_2$ at 0 GPa. With an effective U-J=3 eV on the $d$ orbitals of Mn, the nearest-neighboring Mn-Mn pair, which strongly favors antiferromagnetic coupling without U, takes an ferromagnetic coupling instead. Moreover, the ferromagnetic coupling of the 0-2 Mn-Mn pair is enhanced by about 8 times. Thus, the U term can strongly reduce the antiferromagnetic superexchange of the nearest-neighboring Mn-Mn pair bridged by P atom. We also find that the average spin-polarization on P atoms with considering U is larger than the one without it, indicating enhanced spin polarization of hole carriers. Nevertheless, the previous studies on (Ba$_\text{0.75}$K$_\text{0.25}$)(Zn$_\text{0.95}$Mn$_\text{0.05}$)$_2$As$_2$~\cite{NatureCommunications_I, PRB_V} show that the 0-1 Mn-Mn pair takes AFM coupling. As its counterpart, the 0-1 Mn-Mn pair in (Ba$_\text{0.75}$K$_\text{0.25}$)(Zn$_\text{0.944}$Mn$_\text{0.056}$)$_2$P$_2$ should also be with AFM coupling, which contradicts with the GGA+U results but instead consists with the GGA ones. So the correlation effect in this BaZn$_\text{2}$P$_\text{2}$-based compound is unimportant.

\section{Discussion}
\label{sec:discussion}

We implement further first-principles calculations on the (Ba$_\text{0.75}$K$_\text{0.25}$)(Zn$_\text{0.944}$Mn$_\text{0.056}$)$_2$As$_2$ compound at the GGA level, in order to make a comparison with the (Ba$_\text{0.75}$K$_\text{0.25}$)(Zn$_\text{0.944}$Mn$_\text{0.056}$)$_2$P$_2$ compound. At ambient pressure, the calculated local moments on Mn atoms for all the defined Mn-Mn pairs [Fig. \ref{fig:Structure}(b)] in (Ba$_\text{0.75}$K$_\text{0.25}$)(Zn$_\text{0.944}$Mn$_\text{0.056}$)$_2$P$_2$ range from 3.25 to 3.50 $\mu_{B}$, which are about 0.2 $\mu_{B}$ smaller than those in (Ba$_\text{0.75}$K$_\text{0.25}$)(Zn$_\text{0.944}$Mn$_\text{0.056}$)$_2$As$_2$. This is a signal that the $p$-$d$ hybridization is stronger in the former one. Furthermore, when two Mn atoms are located at different Zn layers in the supercell, the ferromagnetic couplings for all Mn-Mn pairs in (Ba$_\text{0.75}$K$_\text{0.25}$)(Zn$_\text{0.944}$Mn$_\text{0.056}$)$_2$P$_2$ are stronger than those in (Ba$_\text{0.75}$K$_\text{0.25}$)(Zn$_\text{0.944}$Mn$_\text{0.056}$)$_2$As$_2$ (Table~\ref{tab:I}). The stronger effective FM couplings between interlayer Mn-Mn pairs in (Ba$_\text{0.75}$K$_\text{0.25}$)(Zn$_\text{0.944}$Mn$_\text{0.056}$)$_2$P$_2$ may partially help the improvement of the $T_{c}$.

Nevertheless, in both compounds, the 0-1 Mn-Mn pair takes robust AFM coupling, which reduces the concentration of the effective FM coupling of Mn$^{2+}$ ions and leads to the reduction of net magnetization~\cite{NatureCommunications_I, PRB_V}. Moreover, among all configurations for (Ba$_\text{0.75}$K$_\text{0.25}$)(Zn$_\text{0.944}$Mn$_\text{0.056}$)$_2$P$_2$, the AFM coupled 0-1 one has the lowest energy, indicating that the Mn atoms would like to form clusters under an equilibrium growth condition. This goes against the improvement of the sample quality and the $T_{c}$.

Based on above analyses, the substitution of P for As can modulate the magnetic coupling effectively, especially the strength of effective FM coupling between the interlayer Mn-Mn pairs. However, the robust AFM superexchange always induces the AFM coupling for the nearest-neighboring Mn-Mn pairs in both compounds, and may hinder the improvement of sample quality and $T_{c}$ for this II-II-V type DMS materials. Therefore, \textit{searching for a material with intrinsic $p$-$d$ hybridization and transition metal ions far apart without direct bridging anions may serve as a new clue to exploring more feasible DMS materials}.

\section{Conclusions}
\label{sec:summary}

We have systematically studied the magnetic interactions in the proposed BaZn$_\text{2}$P$_\text{2}$-based diluted magnetic semiconductors by using first-principles electronic structure calculations. For the compound with only spin doping, the antiferromagnetic coupling by the short-range AFM superexchange dominates, while the distant Mn-Mn pairs do not show apparently favored magnetic coupling due to the weak interactions. For the compound with both spin and hole dopings, except for several very near Mn-Mn pairs, the ferromagnetic coupling prevails. This originates from the combined effects of the $p$-$d$ exchange between the Mn $d$ orbitals and the neighboring P $p$ orbitals as well as the long-distance interactions transmitted by spin-polarized itinerant hole carriers, while the latter is indeed a critical factor for the rising of ferromagnetism. Furthermore, with applied pressure, the ferromagnetism in (Ba$_\text{0.75}$K$_\text{0.25}$)(Zn$_\text{0.944}$Mn$_\text{0.056}$)$_\text{2}$P$_\text{2}$ is first strengthened and then weakened due to the competition among the pressure-induced changes in the $p$-$d$ hybridization, the band broadening, and the spin polarization of itinerant carriers. The robust AFM coupling of the short-range Mn-Mn pairs bridged by anions hinders the improvement of the sample quality and the $T_{c}$ for this II-II-V type DMSs . We propose that the combination of intrinsic $p$-$d$ hybridization and far apart magnetic ions may be a new clue to searching for more feasible DMS materials.

\begin{acknowledgments}
We thank C. Q. Jin, F. Sun, B. J. Chen, and D. Haskel for helpful discussions. This work was supported by National Key R\&D Program of China (Grant No. 2017YFA0302903), National Natural Science Foundation of China (Grants No. 11774422 and No. 11774424), the Fundamental Research Funds for the Central Universities, and the Research Funds of Renmin University of China (Grants No. 14XNLQ03 and No. 16XNLQ01). Computational resources were provided by the Physical Laboratory of High Performance Computing at RUC.
\end{acknowledgments}


\begin{thebibliography}{}

\bibitem{Science_I} Ohno H \href{http://science.sciencemag.org/content/281/5379/951}{1998 Science \textbf{281} 951}

\bibitem{Science_II} Wolf S A, Awschalom D D, Buhrman R A, Daughton J M, von Moln{\'a}r S, Roukes M L, Chtchelkanova A Y, and Treger D M \href{http://science.sciencemag.org/content/294/5546/1488}{2001 Science \textbf{294} 1488}

\bibitem{Nature_I} Ohno H, Chiba D, Matsukura F, Omiya T, Abe E, Dietl T, Ohno Y, and Ohtani K \href{http://dx.doi.org/10.1038/35050040}{2000 Nature \textbf{408} 944}

\bibitem{NatureMaterials_I} Dietl T \href{http://dx.doi.org/10.1038/nmat2898}{2010 Nat. Mater. \textbf{9} 965}

\bibitem{RevModPhys_I} \v{Z}uti\'{c} I, Fabian J, and Das Sarma S \href{http://link.aps.org/doi/10.1103/RevModPhys.76.323}{2004 Rev. Mod. Phys. \textbf{76} 323}

\bibitem{ZJH} Zhao J H \href{http://dx.doi.org/10.1360/N972015-01392}{2016 Chin. Sci. Bull. \textbf{61} 1401}

\bibitem{ZJH_I} Pan D, Wang S L, Wang H L, Yu X Z, Wang X L, and Zhao J H \href{http://dx.doi.org/10.1088/0256-307X/31/7/078103}{2014 Chin. Phys. Lett. \textbf{31} 078103}

\bibitem{NatureMaterials_II} Sharma P, Gupta A, Rao K V, Owens F J, Sharma R, Ahuja R, Osorio Guillen J M, Johansson B, and Gehring G A, \href{http://dx.doi.org/10.1038/nmat984}{2003 Nat. Mater. \textbf{2} 673}

\bibitem{PRL_I} Saito H, Zayets V, Yamagata S, and Ando K \href{http://link.aps.org/doi/10.1103/PhysRevLett.90.207202}{2003 Phys. Rev. Lett. \textbf{90} 207202}

\bibitem{APL_I} Ohno H, Shen A, Matsukura F, Oiwa A, Endo A, Katsumoto S, and Iye Y \href{http://scitation.aip.org/content/aip/journal/apl/69/3/10.1063/1.118061}{1996 Appl. Phys. Lett. \textbf{69} 363}

\bibitem{APL_II} Overberg M E, Gila B P, Abernathy C R, Pearton S J, Theodoropoulou N A, McCarthy K T, Arnason S B, and Hebard A F \href{http://scitation.aip.org/content/aip/journal/apl/79/19/10.1063/1.1416472}{2001 Appl. Phys. Lett. \textbf{79} 3128}

\bibitem{PRB_I} Soo Y L, Huang S W, Ming Z H, Kao Y H, Munekata H, and Chang L L \href{http://link.aps.org/doi/10.1103/PhysRevB.53.4905}{1996 Phys. Rev. B \textbf{53} 4905}

\bibitem{Nano_I} Chen L, Yang X, Yang F H, Zhao J H, Misuraca J, Xiong P, and von Molnš¢r S \href{https://doi.org/10.1021/nl201187m}{2011 Nano. Lett. \textbf{11} 2584}

\bibitem{JAP_I} Furdyna J K \href{http://scitation.aip.org/content/aip/journal/jap/64/4/10.1063/1.341700}{1988 J. Appl. Phys. \textbf{64} R29}

\bibitem{PRB_II} Kudrnovsk\'{y} J, Turek I, Drchal V, M\'{a}ca F, Weinberger P, and Bruno P \href{http://link.aps.org/doi/10.1103/PhysRevB.69.115208}{2004 Phys. Rev. B \textbf{69} 115208}

\bibitem{PRB_III} Dietl T, Haury A, and Merle d'Aubign\'{e} Y \href{http://link.aps.org/doi/10.1103/PhysRevB.55.R3347}{1997 Phys. Rev. B \textbf{55} R3347}

\bibitem{RevModPhys_II} Jungwirth T, Sinova J, Ma\v{s}ek J, Ku\v{c}era J, and MacDonald A H \href{http://link.aps.org/doi/10.1103/RevModPhys.78.809}{2006 Rev. Mod. Phys. \textbf{78}, 809}

\bibitem{RevModPhys_III} Sato K, Bergqvist L, Kudrnovsk\'{y} J, Dederichs P H,  Eriksson O, Turek I, Sanyal B, Bouzerar G, Katayama-Yoshida H, Dinh V A, Fukushima T, Kizaki H, and Zeller R \href{http://link.aps.org/doi/10.1103/RevModPhys.82.1633}{2010 Rev. Mod. Phys. \textbf{82}, 1633}

\bibitem{APL_IV} Baron T, Tatarenko S, Saminadayar K, Magnea N, and Fontenille J \href{http://scitation.aip.org/content/aip/journal/apl/65/10/10.1063/1.112096}{1994 Appl. Phys. Lett. \textbf{65} 1284}

\bibitem{PRL_II} Haury A, Wasiela A, Arnoult A, Cibert J, Tatarenko S, Dietl T, and Merle d'Aubign\'{e} Y \href{http://link.aps.org/doi/10.1103/PhysRevLett.79.511}{1997 Phys. Rev. Lett. \textbf{79} 511}

\bibitem{PRB_IV} Ferrand D, Cibert J, Wasiela A, Bourgognon C, Tatarenko S, Fishman G, Andrearczyk T, Jaroszy\'{n}ski J, Kole\'{s}nik S, Dietl T, Barbara B, and Dufeu D \href{http://link.aps.org/doi/10.1103/PhysRevB.63.085201}{2001 Phys. Rev. B \textbf{63} 085201}

\bibitem{Handbook_I} Matsukura F, Ohno H, and Dietl T \href{http://www.sciencedirect.com/science/article/pii/S1567271909600056}{2002 Handbook of Magnetic Materials \textbf{14} p1-87}


\bibitem{NatureCommunications_Deng} Deng Z, Jin C Q, Liu Q Q, Wang X C, Zhu J L, Feng S M, Chen L C, Yu R C, Arguello C, Goko T, Ning F L, Zhang J S, Wang Y Y,  Aczel A A, Munsie T, Williams T J, Luke G M, Kakeshita T, Uchida S, Higemoto W, Ito T U, Gu B, Maekawa S, Morris G D, and Uemura Y J,  \href{http://doi.org/10.1038/ncomms1425}{2011 Nat. Commun. \textbf{2} 422}

\bibitem{NatureCommunications_I} Zhao K, Deng Z, Wang X C, Han W, Zhu J L, Li X, Liu Q Q, Yu R C, Goko T, Frandsen B, Liu L, Ning F, Uemura Y J, Dabkowska H, Luke G M, Luetkens H, Morenzoni E, Dunsiger S R, Senyshyn A, B\"{o}ni P, and Jin C Q \href{http://www.nature.com/ncomms/journal/v4/n2/suppinfo/ncomms2447_S1.html}{2013 Nat. Commun. \textbf{4} 1442}

\bibitem{PRB_Ning} Ding C, Man H Y, Qin C, Lu J C, Sun Y L, Wang Q, Yu B Q, Feng C M, Goko T, Arguello C J, Liu L, Frandsen B A, Uemura Y J, Wang H D, Luetkens H, Morenzoni E, Han W, Jin C Q, Munsie T, Williams T J, D'Ortenzio R M, Medina T, Luke G M, Imai T, and Ning F L \href{https://doi.org/10.1103/PhysRevB.88.041102}{2013 Phys. Rev. B \textbf{88} 041102(R)}

\bibitem{SSC} Wang X C, Liu Q Q, Lv Y X, Gao W B, Yang L X, Yu R C, Li F Y, and Jin C Q \href{http://dx.doi.org/10.1016/j.ssc.2008.09.057}{2008 Solid State Commun. \textbf{148} 538}

\bibitem{PRL_III} Rotter M, Tegel M, and Johrendt D \href{http://link.aps.org/doi/10.1103/PhysRevLett.101.107006}{2008 Phys. Rev. Lett. \textbf{101} 107006}

\bibitem{JACS} Kamihara Y, Watanabe T, Hirano M, and Hosono H \href{http://doi.org/10.1021/ja800073m}{2008 J. Am. Chem. Soc. \textbf{130} 3296}

\bibitem{CSB} Zhao K, Chen B, Zhao G Q, Li X, Yuan Z, Deng Z, Liu Q Q, and Jin C Q \href{http://doi.org/10.1007/s11434-014-0398-z}{2014 Chin. Sci. Bull. \textbf{59} 2524}

\bibitem{PRB_V} Glasbrenner J K, \v{Z}uti\'{c} I, and Mazin I I \href{http://link.aps.org/doi/10.1103/PhysRevB.90.140403}{2014 Phys. Rev. B \textbf{90} 140403(R)}

\bibitem{ComputMaterSicence_I} Tao H L, Lin L, Zhang Z H, He M, and Song B \href{http://www.sciencedirect.com/science/article/pii/S0927025614006922}{2015 Comput. Mater. Sci. \textbf{98} 93}

\bibitem{SolidStatePhysics_I} Yang J T, Luo S J, and Xiong Y C \href{http://www.sciencedirect.com/science/article/pii/S1293255815001417}{2015 Solid State Sci. \textbf{46} 102}

\bibitem{PRB_VI} Suzuki H, Zhao K, Shibata G, Takahashi Y, Sakamoto S, Yoshimatsu K, Chen B J, Kumigashira H, Chang F-H, Lin H-J, Huang D J, Chen C T, Gu B, Maekawa S, Uemura Y J, Jin C Q, and Fujimori A \href{http://link.aps.org/doi/10.1103/PhysRevB.91.140401}{2015 Phys. Rev. B \textbf{91} 140401(R)}

\bibitem{PAW_I} Bl\"{o}chl P E \href{http://link.aps.org/doi/10.1103/PhysRevB.50.17953}{1994 Phys. Rev. B \textbf{50} 17953}

\bibitem{PAW_II} Kresse G and Joubert D \href{http://link.aps.org/doi/10.1103/PhysRevB.59.1758}{1999 Phys. Rev. B \textbf{59} 1758}

\bibitem{VASP_I} Kresse G and Hafner J \href{http://link.aps.org/doi/10.1103/PhysRevB.47.558}{1993 Phys. Rev. B \textbf{47} 558}

\bibitem{VASP_II} Kresse G and Furthm\"{u}ller J \href{http://www.sciencedirect.com/science/article/pii/0927025696000080}{1996 Comput. Mater. Sci. \textbf{6} 15}

\bibitem{VASP_III} Kresse G and Furthm\"{u}ller J \href{http://link.aps.org/doi/10.1103/PhysRevB.54.11169}{1996 Phys. Rev. B \textbf{54} 11169}

\bibitem{PBE} Perdew J P, Burke K, and Ernzerhof M \href{http://link.aps.org/doi/10.1103/PhysRevLett.77.3865}{1996 Phys. Rev. Lett. \textbf{77} 3865}

\bibitem{DFTU-II} Dudarev S L, Botton G A, Savrasov S Y, Humphreys C J, and Sutton A P \href{https://doi.org/10.1103/PhysRevB.57.1505}{1998 Phys. Rev. B \textbf{57} 1505}

\bibitem{SSC_177} Tao H L, Zhang Z H, Pan L L, He M, and Song B \href{http://dx.doi.org/10.1016/j.ssc.2013.10.010}{2014 Solid State Commun. \textbf{177} 113}

\bibitem{PRL_98} Ma\v{s}ek J, Kudrnovsk\'{y} J, M\'{a}ca F, Gallagher B L, Campion R P, Gregory D H, and Jungwirth T \href{https://doi.org/10.1103/PhysRevLett.98.067202}{2007 Phys. Rev. Lett. \textbf{98} 067202}

\bibitem{ThCrSi} Hellmann A, L\"{o}hken A, Wurth A, and Mewis A \href{http://www.degruyter.com/view/j/znb.2007.62.issue-2/znb-2007-0203/znb-2007-0203.xml}{2007 Z. Naturforsch. B \textbf{62} 155}

\bibitem{JACS_I} Xiao Z, Hiramatsu H, Ueda S, Toda Y, Ran F-Y, Guo J, Lei H, Matsuishi S, Hosono H, and Kamiya T \href{http://pubs.acs.org/doi/abs/10.1021/ja507890u}{2014 J. Am. Chem. Soc. \textbf{136} 14959}

\bibitem{BZNP} Kl\"{u}fers P and Mewis A \href{https://doi.org/10.1515/znb-1978-0207}{1978 Z. Naturforsch. \textbf{33} 151}

\bibitem{JAC_583} Shein I R and Ivanovskii A L \href{http://dx.doi.org/10.1016/j.jallcom.2013.08.118}{2014 J. Alloys Compd. \textbf{583} 100}

\bibitem{PhysicReview_I} Anderson P W \href{http://link.aps.org/doi/10.1103/PhysRev.79.350}{1950 Phys. Rev. \textbf{79} 350}


\bibitem{PhysicReview_II} Zener C \href{http://link.aps.org/doi/10.1103/PhysRev.81.440}{1951 Phys. Rev. \textbf{81} 440}

\bibitem{PRB_VII} Sato K, Schweika W, Dederichs P H, and KatayamaYoshida H \href{http://link.aps.org/doi/10.1103/PhysRevB.70.201202}{2004 Phys. Rev. B \textbf{70} 201202}


\bibitem{PRB_Sun} Sun F, Li N N, Chen B J, Jia Y T, Zhang L J, Li W M, Zhao G Q, Xing L Y, Fabbris G, Wang Y G, Deng Z, Uemura Y J, Mao H K, Haskel D, Yang W G, and Jin C Q \href{https://doi.org/10.1103/PhysRevB.93.224403}{2016 Phys. Rev. B \textbf{93} 224403}

\bibitem{PRB_SunI} Sun F, Zhao G Q, Escanhoela C A, Jr, Chen B J, Kou R H, Wang Y G, Xiao Y. M, Chow P, Mao H K, Haskel D, Yang W G, and Jin C Q \href{https://journals.aps.org/prb/abstract/10.1103/PhysRevB.95.094412}{2017 Phys. Rev. B \textbf{95} 094412}

\bibitem{TuZhu_I} Here, the differential charge density is defined as: C$_\text{AB}$-C$_\text{A}$-C$_\text{B}$. C$_\text{AB}$ denotes the charge density of the supercell with all atoms Ba$_\text{18}$Zn$_\text{34}$Mn$_\text{2}$P$_\text{36}$ ((Ba$_\text{0.75}$K$_\text{0.25}$)$_\text{18}$Zn$_\text{34}$Mn$_\text{2}$P$_\text{36}$), C$_\text{A}$ denotes the charge density of the supercell without two Mn atoms Ba$_\text{18}$Zn$_\text{34}$Mn$_\text{0}$P$_\text{36}$  ((Ba$_\text{0.75}$K$_\text{0.25}$)$_\text{18}$Zn$_\text{34}$Mn$_\text{0}$P$_\text{36}$) and C$_\text{B}$ denotes the charge density of the supercell with only two Mn atoms Ba$_\text{0}$Zn$_\text{0}$Mn$_\text{2}$P$_\text{0}$ ((Ba$_\text{0.75}$K$_\text{0.25}$)$_\text{0}$Zn$_\text{0}$Mn$_\text{2}$P$_\text{0}$).


\end{thebibliography}
\end{document}